\documentclass[aps,twocolumn,secnumarabic,balancelastpage,amsmath,amssymb,nofootinbib,floatfix]{revtex4-1}

\usepackage{graphicx}  
\usepackage{bm}          
\usepackage{tabularx}
\usepackage{lipsum}
\usepackage{gensymb}
\usepackage{multirow}
\usepackage{subfigure}
\usepackage[section]{placeins}
\usepackage[colorlinks=true]{hyperref}

\begin{document}
\title{Effect of Mars Atmospheric Loss on Snow Melt Potential in a 3.5-Gyr Mars Climate Evolution Model}
\author{Megan Mansfield}
\email{meganmansfield@uchicago.edu}
\date{\today}
\affiliation{University of Chicago Department of Geophysical Sciences}

\author{Edwin S. Kite}
\affiliation{University of Chicago Department of Geophysical Sciences}

\author{Michael A. Mischna}
\affiliation{Jet Propulsion Laboratory, California Institute of Technology}

\begin{abstract}
Post-Noachian Martian paleochannels indicate the existence of liquid water on the surface of Mars after about 3.5~Gya \citep{Irwin2015,Palucis2016}. In order to explore the effects of variations in CO$_{2}$ partial pressure and obliquity on the possibility of surface water, we created a zero-dimensional surface energy balance model. We combine this model with physically consistent orbital histories to track conditions over the last 3.5~Gyr of Martian history. We find that melting is allowed for atmospheric pressures corresponding to exponential loss rates of $dP/dt \propto t^{-3.73}$ or faster, but this rate is within $0.5\sigma$ of the rate calculated from initial measurements made by the Mars Atmosphere and Volatile EvolutioN (MAVEN) mission, if we assume all the escaping oxygen measured by MAVEN comes from atmospheric CO$_{2}$ \citep{Lillis2017,Tu2015}. Melting at this loss rate matches selected key geologic constraints on the formation of Hesperian river networks, assuming optimal melt conditions during the warmest part of each Mars year \citep{Irwin2015,Stopar2006,Kite2017b,Kite2017a}. The atmospheric pressure has a larger effect on the surface energy than changes in Mars's mean obliquity. These results show that initial measurements of atmosphere loss by MAVEN are consistent with atmospheric loss being the dominant process that switched Mars from a melt-permitting to a melt-absent climate \citep{Jakosky2017}, but non-CO$_{2}$ warming will be required if $<2$~Gya paleochannels are confirmed, or if most of the escaping oxygen measured by MAVEN comes from H$_{2}$O.
\end{abstract}

\maketitle


\section{Introduction}

Large ($>10 \text{ km}^{2}$), late-stage Martian alluvial fans and river deltas provide evidence for surface liquid water on post-Noachian Mars after about 3.5~Gya \citep{Grant2012,Williams2014,Irwin2015,Palucis2016}. However, by about 3.5~Gya most of the conditions favorable to the existence of surface liquid water no longer existed: much of Mars's atmosphere was lost before the end of the Noachian and the Martian dynamo shut down around the mid-Noachian \citep{Pepin1994,Lillis2013}. Additionally, the amount of greenhouse gases released by volcanism on Mars is less than on Earth because the mantle is more reducing \citep{Stanley2011}, and by 3.5~Gya the rate of volcanic degassing had slowed down significantly \citep{Kite2009}. Because orbiters detect only minor post-Noachian carbonate, which may be due to SO$_{2}$ or acidity \citep{Bullock2007,Halevy2009b}, it is difficult to justify post-Noachian carbonate sequestration of more CO$_{2}$ than exists in the present, thin atmosphere \citep{Edwards2015}.

Atmospheric pressure has a strong effect on permitting surface liquid water because lower atmospheric pressure can preclude melting by decreasing the strength of greenhouse warming and increasing the amount of evaporative cooling. \citet{Clow1987} showed these effects in a model which determined the minimum atmospheric pressure necessary to melt dusty snow on the surface of Mars. They found that melting could occur at relatively low pressures if the snowpack was assumed to be thick enough that melt accumulated at the base of the thick snow layer while the upper layers of snow were at colder temperatures. \citet{Hecht2002} used a similar model to determine the necessary surface properties, such as albedo and conductivity, to allow melting on Mars. Additionally, Hecht emphasized the importance of evaporative cooling at low atmospheric pressures, as first pointed out by \citet{Ingersoll1970}. Models which have not included evaporative cooling produce much more melting on the surface of Mars, which indicates its importance in setting melting conditions \citep{Costard2002,WilliamsK2008}. The high rate of evaporative cooling at low atmospheric pressure precludes melting of pure water ice on present-day Mars, which has an average atmospheric pressure of 600~Pa.

Although atmospheric pressure can have a strong effect on the surface energy balance, changes in Mars's orbital parameters, especially large changes in its mean obliquity, can also affect melt conditions. \citet{Kite2013} determined the conditions for snowmelt at a variety of atmospheric pressures, obliquities, and eccentricities. They found that combining these factors could produce intermittent melting that could provide small amounts of liquid water for sediment induration at low latitudes.

While models that study moments in time can provide some insight into the past Martian climate, models showing evolution over time give a better understanding of the timing of melting in Mars's past. \citet{Manning2006} addressed the time evolution of the Martian climate by examining the relative sizes of various CO$_{2}$ reservoirs over the history of Mars. Their model determined that different climate states are stable for different obliquities, but they only considered a subset of the full range of obliquities that Mars has experienced.

A model that simultaneously studies the effects of time evolution, orbital variations, and changing energy balance with changing atmospheric pressure is needed to more fully understand the timing and intermittency of melting on Mars. To address this, we created a zero-dimensional energy balance model that spans over 3.5~Gyr of Mars history and examines how changes in orbital parameters, solar luminosity, and atmospheric pressure impact the melt conditions of snow on the surface of Mars. We consider snowmelt as a source of water as opposed to catastrophic flows after impacts because recent work suggests lakes lasted for at least a few thousand years, and this relatively long lifetime implies a climate favorable to melting and not just a sudden catastrophic event that allowed melting for a brief period \citep{Grant2012,Williams2014,Irwin2015,Palucis2016,Turbet2017}. We also ignore the effects of impacts and eruptions because changes due to atmospheric loss, solar brightening, and orbital variations are more well understood, and because the melt potential of impacts themselves depends on the atmospheric heat capacity and therefore the atmospheric pressure. Our model is zero-dimensional because this allows fast computation of a full 3.5-Gyr energy balance, whereas previous 3D models have been unable to include time evolution because of computational limits \citep{Wordsworth2013}. This model is timely because the Mars Atmosphere and Volatile EvolutioN (MAVEN) mission is currently in orbit around Mars collecting data on rates of atmospheric loss, and is presumably providing better constraints on the atmospheric pressure over time on Mars which we can use in our model to constrain the conditions that explain melting on post-Noachian Mars \citep{Lillis2017}. We describe our model in Section \ref{sec:model}. In Section \ref{sec:results} we present our results, which indicate that atmospheric pressure has a more dominant effect on surface melt conditions than obliquity variations. In Section \ref{sec:discussion} we discuss assumptions of our model and possible future extensions, and we summarize our findings in Section \ref{sec:conclusions}.

\section{Model Description}
\label{sec:model}

\subsection{Chaotic Orbital Histories}
\label{sec:orbits}

\begin{figure}[h]
\centering
\includegraphics[width=\linewidth]{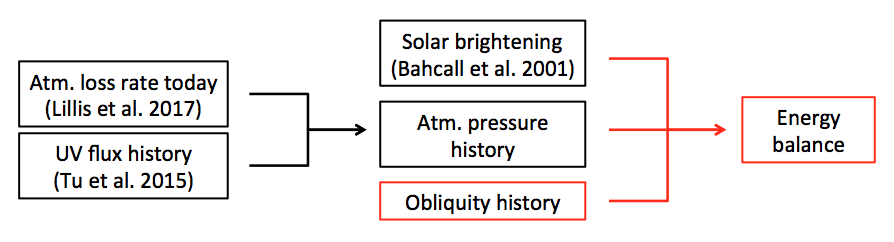}
\caption{\label{fig:flowchart} A flowchart showing how key aspects of the model were incorporated. Boxes outlined in red indicate new contributions from our model. Atmospheric loss, solar brightening, and obliquity histories were calculated as described in Section \ref{sec:orbits} and combined in an energy balance described in section \ref{sec:inmodel}.}
\end{figure}

\begin{figure}
\centering
	\begin{subfigure}{}
	\includegraphics[width=\linewidth]{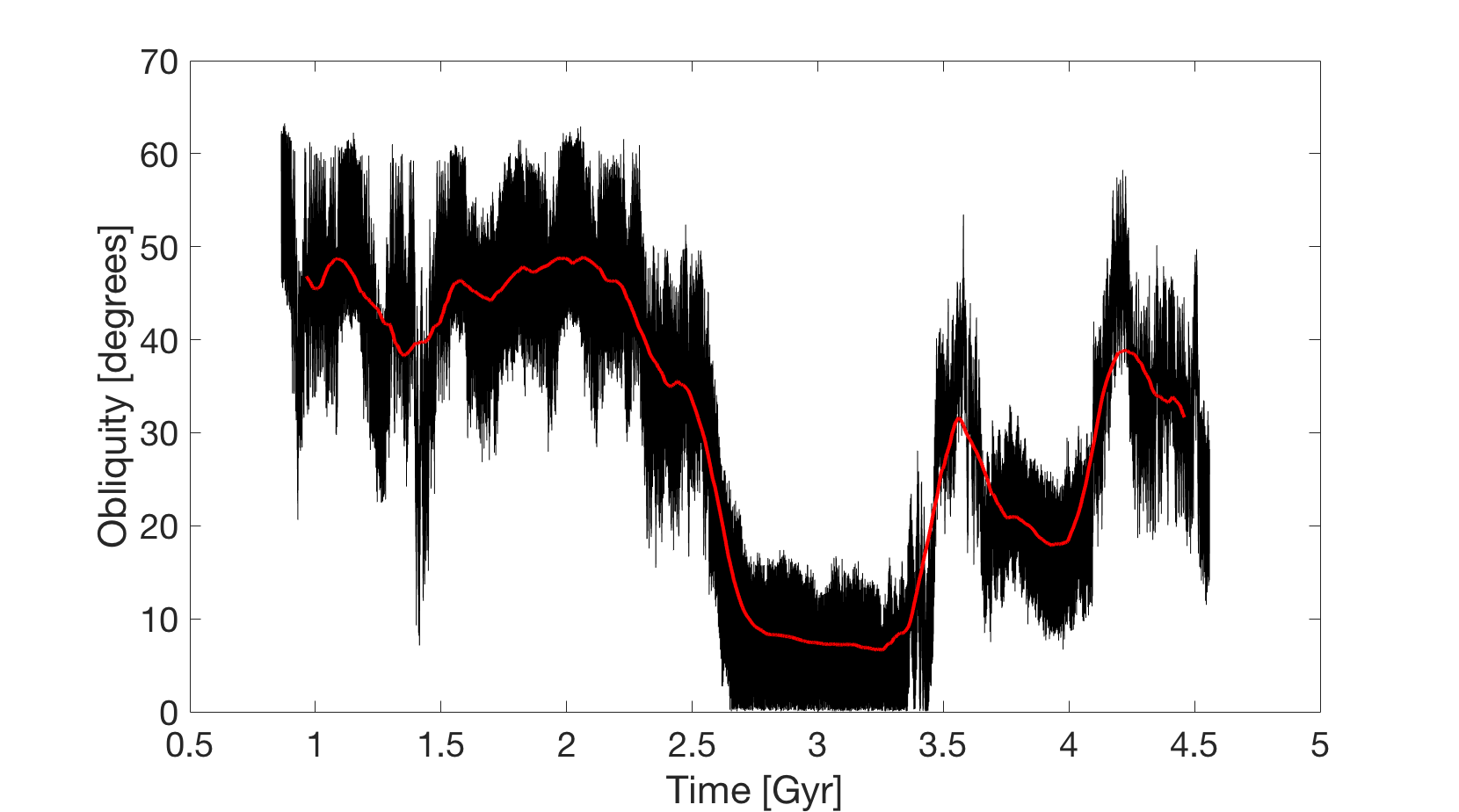}
	\end{subfigure}
	\begin{subfigure}{}
	\includegraphics[width=\linewidth]{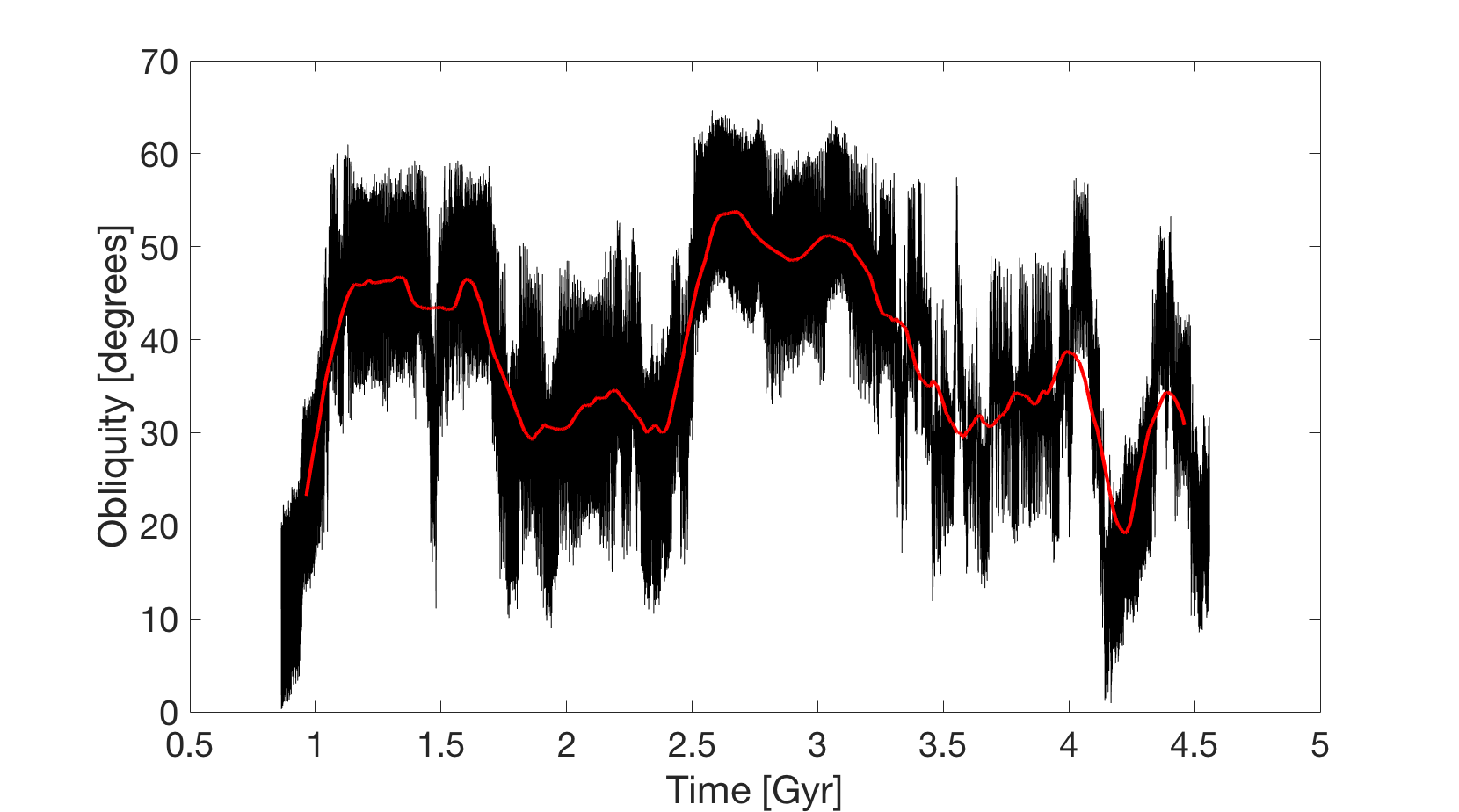}
	\end{subfigure}
\caption{\label{fig:obl23} Obliquity histories for orbital tracks 1 (top) and 2 (bottom). The overplotted red lines show 200-Myr averages.}
\end{figure}

To investigate the influence of orbital variability on post-Noachian surface liquid water, we created a model that combined the effects of orbital variations, atmospheric loss, and solar brightening, as illustrated in Figure \ref{fig:flowchart}. Physically consistent orbital histories were constructed using the {\fontfamily{bch}\selectfont mercury6} N-body integrator and an obliquity code \citep{Chambers1999,Armstrong2014}. The orbital histories recorded the orbital parameters of the simulated Mars at 200-year intervals over the last 3.5~Gyr and included quasi-periodic variations in orbital parameters \citep{Kite2015}. Most importantly, they included changes in obliquity, which for Mars has probably jumped from low mean values around $20\degree$ to high mean values around $40\degree$  between two and nine times in its history on $\gtrapprox$ 200-Myr timescales \citep{Li2014}. The orbital histories were sampled at 2,000-year intervals. Eight possible orbital histories were studied, all of which have obliquity jumps at different times in their timelines, but which all end with obliquities within 12\degree~of the modern-day value of $25.19\degree$. Several obliquity tracks were considered because the chaotic nature of the Solar System means that the exact value of Mars's obliquity is not known beyond a few hundred million years ago. Therefore, we studied a distribution of possible orbital histories to statistically constrain the influence of the orbital parameters on melting. The eight orbital histories all produced similar energy histories, so only two will be examined in detail in this paper. Obliquity histories for tracks 1 and 2, which will be examined in more detail later, are shown in Figure \ref{fig:obl23}.

\subsection{Loss of Atmospheric Pressure}

The chaotic orbital histories were combined with a model of solar brightening over time and loss of atmospheric pressure \citep{Bahcall2001}. Although there are several possible loss mechanisms for CO$_{2}$ in past Mars, such as chemical fixation of CO$_{2}$ in deep aquifers \citep{Chassefiere2011} and basal melting of a CO$_{2}$ ice cap \citep{Kurahashi2006}, we chose to focus on escape to space and assumed all atmospheric loss was due to escape to space.
 Additionally, we neglected carbonate formation because it is thought to be relatively ineffective in the post-Noachian, although this is debated \citep{Edwards2015,Hu2015}. 
 
 The current atmospheric pressure on Mars is approximately 600~Pa, but at the temperatures that would allow liquid water on the surface of Mars, the approximately 600~Pa of additional CO$_{2}$ currently trapped in Martian polar caps would be released into the atmosphere, and so we assume a modern-day atmospheric pressure of 1200~Pa \citep{Bierson2016}. Atmospheric loss to space was parameterized as a power law
\begin{equation}
\label{eq:atmloss}
\frac{dP}{dt}=-k\left( \frac{t_{0}}{t} \right)^{\alpha}
\end{equation}
where $P$ is the atmospheric pressure in Pa, $t$ is time in Gyr since the Sun formed, $t_{0}=4.56 \text{ Gyr}$, and $k=1270 \text{ Pa Gyr}^{-1}$ is a constant chosen to match the parameterization to the current rate of Mars atmospheric loss due to solar UV flux, which we estimated MAVEN measurements of hot atomic oxygen escape \citep{Lillis2017}. This estimate of $k$ assumes that the oxygen loss which was measured by MAVEN corresponds entirely to loss of CO$_{2}$ over geologic time, and not to loss of other atmospheric constituents such as H$_{2}$O. We assume that the oxygen loss is due to loss of CO$_{2}$ because Mars does not have significant carbonate deposits and a thick CO$_{2}$ atmosphere was likely necessary on early Mars. However, MAVEN has yet to identify a major loss channel for carbon, and some authors \citep[e.g.,][]{Hodges2002} argue that Mars has lost negligible amounts of carbon to space since the Noachian, so that almost all of the oxygen loss corresponds to net loss of H$_{2}$O to space. Therefore, our assumption that all of the oxygen loss is due to loss of CO$_{2}$ is the most optimistic situation for producing higher past atmospheric pressures. $\alpha$ represents the increase in atmospheric loss rate in the past relative to the current loss rate, due to increased solar UV flux earlier in the Sun's lifetime \cite{Tu2015}. In the model, $\alpha$ was varied between $\alpha=0$ and $\alpha=4.24$. A variety of values of $\alpha$ were investigated because, while the solar brightening over time is well understood \citep{Bahcall2001}, the atmospheric pressure over time is poorly understood. A constant loss rate with $\alpha=0$ is unrealistic, as the past atmospheric loss should have been faster because of higher solar UV flux \citep{Tu2015}, flaring, and solar wind, but we include this value as a reference. $\alpha=4.24$ was the maximum value considered because it leads to an atmospheric pressure of approximately 2~bars at 3.5~Gya. 2~bars is estimated to be an approximate upper limit on the atmospheric pressure at about 3.5~Gya from the size distribution of craters formed at this time \citep{Kite2014}. The current best estimate of a realistic value for $\alpha$ comes from an estimate of modern-day atmospheric loss on Mars based on initial MAVEN results, combined with an estimate of the evolution of the Sun's UV flux over time \citep{Lillis2017,Tu2015}. The initial results from the MAVEN mission suggest that the average modern rate of photochemical loss of oxygen is $4.3 \times 10^{25}$~s$^{-1}$, with upper and lower bounds of $9.6 \times 10^{25}$~s$^{-1}$ and $1.9 \times 10^{25}$~s$^{-1}$, respectively. Our estimate considers only photochemical loss of oxygen, but not ion escape or sputtering. (See \citet{Lillis2015} for a review of these fluxes and how they are measured). Currently published estimates indicate that the present-day escape rate due to ion escape is approximately one order of magnitude lower than photochemical loss \citep{Barabash2007,Ramstad2015,Dong2017}, and that the present-day escape rate due to sputtering is smaller still \citep{Leblanc2015}. These escape rates would differ in the past, because the solar wind evolves as the Sun ages. However, in part because of the uncertain rotation history of the Sun and the difficulty of directly observing stellar winds from solar-analog stars, there is still significant uncertainty about how to extrapolate solar wind interaction into the past \citep{Johnstone2015}. Therefore, based on publications to date, it appears that photochemical loss of oxygen is the dominant loss channel for mass loss from today's Mars. If either ion escape or sputtering played a proportionately more important role in the past, then this would have the effect of increasing $\alpha$ in our two-parameter model given in Equation \ref{eq:atmloss}.

Table \ref{tab:maxP} shows the maximum atmospheric pressure in the model (the pressure at 3.5~Gya) for $\alpha=3.22$, which is the mean value estimated by \citet{Lillis2017} using data from MAVEN and an estimate of the history of solar UV flux from \citet{Tu2015}, and values that are 0.5 and 1 standard deviations away from this value. To calculate values of $\alpha$ corresponding to 0.5 and 1 standard deviations away from $\alpha=3.22$, we combined error estimates from \citet{Lillis2017} and \citet{Tu2015}. While \citet{Lillis2017} quote $1\sigma$ errors on their measurement, \citet{Tu2015} only provide information on the 10th, 50th, and 90th percentiles of their calculations, and so we assumed a Gaussian distribution to estimate $0.5\sigma$ and $1\sigma$ errors from \citet{Tu2015}. The ongoing MAVEN mission will continue to improve estimates of $\alpha$ \citep{Lillis2017}.

\begin{table}
\centering
\caption{\label{tab:maxP} Maximum pressure in the 3.5-Gyr energy balance for different rates of atmospheric loss, as given by Equation \ref{eq:atmloss}, assuming that all oxygen loss measured by MAVEN corresponds to loss over geologic time of CO$_{2}$ and not H$_{2}$O \citep{Lillis2017}. $\alpha=3.22$ is the mean value of $\alpha$ inferred from initial MAVEN results, while $\alpha=2.20$, 2.71, 3.73, and 4.24 represent values that are $-1$, $-0.5$, 0.5, and 1 standard deviations away from this value, respectively \citep{Lillis2017,Tu2015}.}
\begin{tabular}{c | | c | c | c | c | c}
$\alpha$ & 2.20 & 2.71 & 3.22 & 3.73 & 4.24 \\
\hline
Maximum Pressure [bar] & 0.243 & 0.390 & 0.654 & 1.133 & 2.020 \\
\end{tabular}
\end{table}

\subsection{Snow Surface Energy Balance Model}
\label{sec:inmodel}

\begin{figure}
\centering
\includegraphics[width=\linewidth]{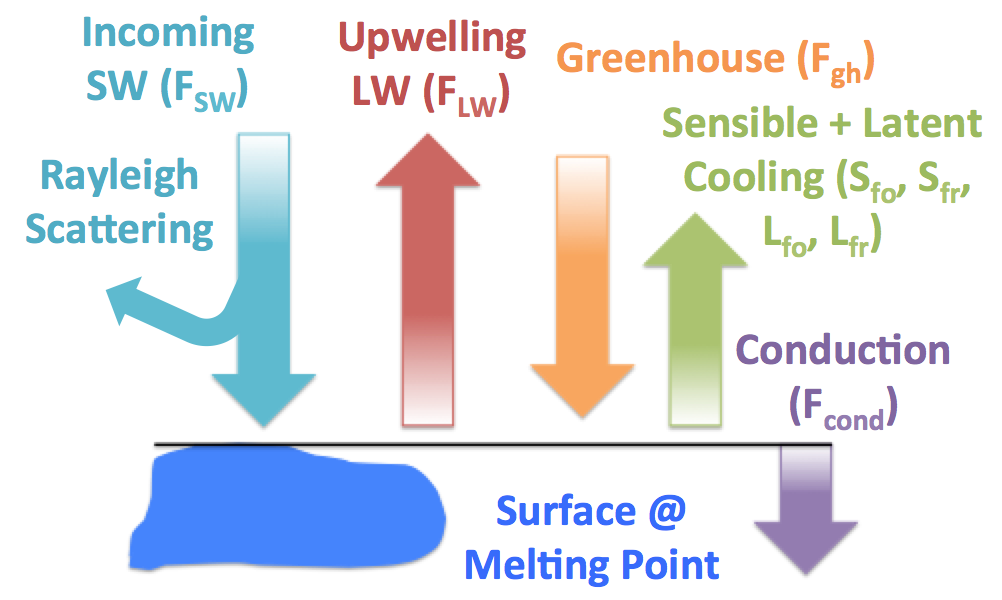}
\caption{\label{fig:fluxcartoon} Schematic of the fluxes included in the energy balance model for a snow surface at T=273 K.}
\end{figure}

To evaluate whether the orbital histories could produce melting conditions, we wrote a zero-dimensional energy balance model in MATLAB for a snow surface on Mars. We primarily considered snow surfaces at low latitudes beween $0\degree$ and $20\degree$ because we expect these latitudes to be the most stable locations for snow and ice when the obliquity is higher than $40\degree$, but we also considered slightly higher latitudes of $40\degree$ at which ice would be stable for intermediate obliquities between $30\degree$ and $40\degree$ \citep{Jakosky1985,Mischna2003,Forget2006}. Because we only consider melting when the obliquity is high, the pole temperature should be high enough to prevent atmospheric collapse and to swiftly reverse a preexisting collapse \citep{Forget2013, Soto2015}. We assumed a snow surface at the melting point ($T=273$~K) and calculated whether the net energy delivered to the snow surface per second was positive, which would indicate melting was possible, or negative, which would indicate that the snow surface would not be warm enough to melt. We assumed an albedo for dusty snow of 0.3, a thermal inertia of $\approx$275~J~m$^{-2}$~K$^{-1}$~s$^{-1/2}$, and a relative humidity of 0.25 \citep{Kite2013}. These assumptions will be discussed more in Section \ref{sec:assume}. The energy balance model incorporated the effects of insolation, upwelling longwave radiation from the surface, the greenhouse effect from the CO$_{2}$ atmosphere, latent cooling due to evaporation, sensible cooling due to atmosphere-surface temperature differences, and conduction. We do not consider the effects of meridional advection. Additionally, we do not include warming from water ice clouds, because the amplitude of warming depends on the optical depth and altitude of the clouds, and these are unknown for early Mars \citep{Ramirez2017}. Figure \ref{fig:fluxcartoon} shows a schematic of the terms in the energy balance.

The incoming solar flux, $F_{SW}$, was estimated for a peak melting period measured over the warmest four hours in the warmest season of the year. We chose to calculate the energy balance only for the most optimal melting conditions in the year because periodic warm conditions could still result in melting for part of the year that could create the observed large alluvial fan features \citep{Clow1987}. In contrast to previous studies \citep[e.g.,][which looked at averaged temperatures for ground ice]{Schorghofer2008}, we only consider peak temperatures because we are interested in studying surface ice and snow. The effect of Rayleigh scattering was included \citep{Kite2013}.

The upwelling longwave radiation, $F_{LW}$, was modeled as a grey body following the equation
\begin{equation}
\label{eq:olr}
F_{LW}=\epsilon \sigma T^{4}
\end{equation}
where $\sigma$ is the Stefan-Boltzmann constant, the emissivity $\epsilon$ was assumed to be 0.98, and the temperature was held fixed at the melting point ($T=273 \text{ K}$). Values for all constants used in the energy balance equations are given in Table \ref{tab:constants}.

\begin{table*}
\centering
\caption{\label{tab:constants} Values of key constants used in the energy balance equations.}
\begin{tabular}{>{\centering \arraybackslash}p{0.5 cm} >{\centering \arraybackslash}p{4.2 cm} >{\centering \arraybackslash}p{3.4 cm} >{\centering \arraybackslash}p{4.2 cm}}
\hline
Symbol & Parameter & Value & Source\\
\hline
$\sigma$ & Stefan-Boltzmann constant & $5.67 \times 10^{-8}$ W m$^{-2}$ K$^{-4}$ & \\
$\epsilon$ & emissivity of ice & 0.98 & \\
$k$ & thermal conductivity of snow & 0.125 W/m/K & Carr and Head (2003) \\
$\tau$ & length of Martian day & 88200 s & \\
$d$ & diurnal skin depth & 0.22 m & Turcotte and Schubert (2014) \\
$T_{s}$ & surface temperature & 273.15 K & \\
$k_{a}$ & thermal conductivity of atmosphere & 0.0138 W/m/K & Vesovic et al. (1990) \\
$C_{p}$ & specific heat capacity of atmosphere & 770 J/kg/K & Kite et al. (2013) \\
$g$ & Mars surface gravity & 3.7 m/s$^{2}$ & \\
$m_{c}$ & molar mass of CO$_{2}$ & 0.044 kg/mol & \\
$m_{w}$ & molar mass of H$_{2}$O & 0.018 kg/mol & \\
$r_{h}$ & atmospheric relative humidity & 0.25 & \\
$A_{v}$ & von Karman's constant & 0.4 & \\
$z_{anem}$ & anemometer height & 5.53 m & Kite et al. (2013) \\
$z_{0}$ & surface roughness & $10^{-4}$ m & Dundas and Byrne (2010) \\
$L_{e}$ & latent heat of evaporation & $2.83 \times 10^{6}$ J/kg & \\
$M_{w}$ & molecular mass of H$_{2}$O & $2.99 \times 10^{-26}$ kg & \\
$k$ & Boltzmann's constant & $1.381 \times 10^{-23}$ J/K & \\
\hline
\end{tabular}
\end{table*}

The greenhouse effect was estimated based on a fit to a Mars GCM \citep{Mischna2012,Mischna2013}. In the GCM, we used the {\fontfamily{bch}\selectfont radtran} model from \citet{Mischna2012} with no water vapor and present-day topography to estimate the greenhouse effect for atmospheric pressures of 6, 60, 600, and 1200~mbar, and for obliquities of $15\degree$, $25\degree$, $35\degree$, $45\degree$, and $60\degree$. The model used present-day values for Mars's eccentricity (0.0935) and longitude of perihelion ($251\degree$). The greenhouse effect from the atmospheric CO$_{2}$ should scale with the longwave flux out of the surface, which is proportional to the temperature to the fourth power. Additionally, for saturated bands such as the 15~$\mu$m CO$_{2}$ absorption band, the greenhouse effect is proportional to the square root of the pressure because of pressure broadening \citep{Goody1989}. Therefore, a linear regression was performed to find a fit to an equation in the form $F_{gh}=a + bT^{4}\sqrt{P}$, where $T$ is the surface temperature in Kelvin, $P$ is the atmospheric pressure in Pa, and $a$ and $b$ are constants. Before performing the fit, points that had elevations lower than $-6$~km or higher than $11$~km, albedo less than 0.15, or thermal inertia less than $150$~J~m$^{-2}$~K$^{-1}$~s$^{-1/2}$ were removed from the model output. Additionally, we considered only points with latitude between $-25 \degree$ and $25 \degree$, because we are primarily interested in melting near the equator and because this resulted in a data set with less scatter \citep{Kraal2008}. The final fit to the greenhouse forcing gave
\begin{equation}
\label{eq:green}
F_{gh}=(9.981 \pm 0.145)+(9.662 \pm 0.017) \times 10^{-9}T^{4}\sqrt{P}
\end{equation}
with $F_{gh}$ in W~m$^{-2}$, $T$ in K, and $P$ in Pa. The fit had R$^{2}=0.98$.

The GCM was also used to estimate a diurnal temperature range, which was used to calculate conductive cooling. The conductive cooling calculated over the same 4-hour peak melting period as the incoming solar flux was calculated using
\begin{equation}
\label{eq:cond}
F_{cond}=\frac{\Delta T k \sin(7200 \omega)}{14400 \omega d}
\end{equation}
where $\Delta T$ is the diurnal temperature range estimated from the Mars GCM \citep{Mischna2012,Mischna2013}, $k$ is the thermal conductivity of the snow surface, $\omega=\frac{2\pi}{\tau}$ s$^{-1}$ is the frequency of temperature oscillations, $\tau$ is the length of a Martian day, and $d$ is the diurnal skin depth.

The forced and free sensible and latent cooling fluxes were calculated following the parameterizations in \citet{Dundas2010}, \citet{Hecht2002}, and \citet{Ingersoll1970}. Free sensible cooling is due to the buoyancy of air near the surface and was described by
\begin{equation}
\label{eq:freesens}
S_{fr}=0.14(T_{s}-T_{a})k_{a}\left[\left(\frac{C_{p}\nu_{a}\rho_{a}}{k_{a}}\right)\left(\frac{g}{\nu_{a}^{2}}\right)\left(\frac{\Delta \rho}{\rho_{a}}\right)\right]^{1/3}
\end{equation}
where $T_{s}$ is the surface temperature, $T_{a}$ is the atmospheric temperature, $k_{a}$ is the atmospheric thermal conductivity, $C_{p}$ is the specific heat capacity of air, $\nu_{a}$ is the air viscosity, $\rho_{a}$ is the density of the atmosphere, $g$ is Mars gravity, and $\frac{\Delta \rho}{\rho_{a}}$ is given by
\begin{equation}
\label{eq:delrho}
\frac{\Delta \rho}{\rho_{a}}=\frac{(m_{c}-m_{w})e_{sat}(1-r_{h})}{m_{c}P}
\end{equation}
where $m_{c}$ is the molar mass of CO$_{2}$, $m_{w}$ is the molar mass of water, $e_{sat}$ is the saturation vapor pressure at temperature $T_{s}$, $r_{h}$ is the relative humidity of the atmosphere, and $P$ is the atmospheric pressure in Pa \citep{Kite2013}. The thermal conductivity $k_{a}$ was determined by interpolating data from \citet{Vesovic1990} to a temperature of $T=257$~K, which was the average atmospheric temperature in our model. The atmospheric temperature $T_{a}$ was determined from a fit to data for the warmest part of the warmest day of the year from the Mars GCM described above, based on the model in \citet{Mischna2012}. These data were used because our model determines the surface energy balance for the warmest part of the year. Atmospheric temperatures were determined from a linear fit of $\ln{T_{a}}$ vs. $\ln{P}$, which had R$^{2}=0.85$. The air viscosity was given by
\begin{equation}
\label{eq:nua}
\nu_{a}=(1.48 \times 10^{-5})\left(\frac{RT_{bl}}{m_{c}P}\right)\left(\frac{240+293.15}{240+T_{bl}}\right)\left(\frac{T_{bl}}{293.15}\right)^{\frac{3}{2}}
\end{equation}
where $R$ is the universal gas constant and $T_{bl}$ is the temperature of the atmospheric boundary layer \citep{Dundas2010}. The boundary layer temperature was determined from a fit to the Mars GCM \citep{Mischna2012} in the same manner as the fit to the atmospheric temperature $T_{a}$. The density of the atmosphere was determined using the ideal gas law.

Forced sensible cooling is due to advection by the wind and was given by
\begin{equation}
\label{eq:forcedsens}
S_{fo}=\rho_{a}C_{p}u_{s}A(T_{s}-T_{a})
\end{equation}
where $u_{s}$ is the wind speed near the surface, and $A$ is given by
\begin{equation}
\label{eq:A}
A=\frac{A_{v}^{2}}{\ln(z_{anem}/z_{0})^{2}}
\end{equation}
where $A_{v}$ is von Karman's constant, $z_{anem}$ is the anemometer height, and $z_{0}$ is the surface roughness \citep{Dundas2010}. The wind speed was also determined from a fit to the Ames Mars GCM wind speeds \citep{Kahre2006}. A linear fit of $\ln{u_{s}}$ vs. $\ln{P}$ for pressures of 7, 50, and 80~mbar was extrapolated to higher pressures. The fit had R$^{2}=0.96$.

Free latent cooling was given by
\begin{equation}
\label{eq:freelat}
L_{fr}=1.57 \times 0.14\, L_{e}\Delta \eta \rho_{a}D_{a}\left[\left(\frac{\nu_{a}}{D_{a}}\right)\left(\frac{g}{\nu_{a}^{2}}\right)\left(\frac{\Delta \rho}{\rho_{a}}\right)\right]^{1/3}
\end{equation}
where $L_{e}$ is the latent heat of evaporation, $\Delta \eta$ is the difference between the water mass fractions in the atmosphere and surface, and $D_{a}$ is the diffusion coefficient of H$_{2}$O in CO$_{2}$ \citep{Dundas2010}. The factor of 1.57 was added to the beginning of the equation to make the amount of latent cooling match experimental results for evaporative losses at current Mars pressure \citep{Moore2006}. $\Delta \eta$ was given by
\begin{equation}
\label{eq:deleta}
\Delta \eta = \frac{\rho_{sat}(1-r_{h})}{\rho_{a}}
\end{equation}
where $\rho_{sat}$ is the saturation vapor density, determined from $e_{sat}$ and the ideal gas law \citep{Kite2013}. The diffusion coefficient of H$_{2}$O in CO$_{2}$ was given by
\begin{equation}
\label{eq:da}
D_{a}=(1.387 \times 10^{-5})\left(\frac{T_{bl}}{273.15}\right)^{\frac{3}{2}}\left(\frac{10^{5}}{P}\right)
\end{equation}
where $P$ is in Pa \citep{Dundas2010}.

Forced latent cooling was given by
\begin{equation}
\label{eq:forcedlat}
L_{fo}=L_{e}\left(\frac{M_{w}}{kT_{bl}}\right)u_{s}A(e_{sat}(1-r_{h}))
\end{equation}
where $M_{w}$ is the molecular mass of water and $k$ is Boltzmann's constant \citep{Dundas2010}. Following \citet{Dundas2010}, \citet{WilliamsR2008}, and \citet{Toon1980}, we sum all of the sensible and latent cooling terms together in our energy balance model instead of considering one dominant term.

\begin{figure}
\centering
\includegraphics[width=\linewidth]{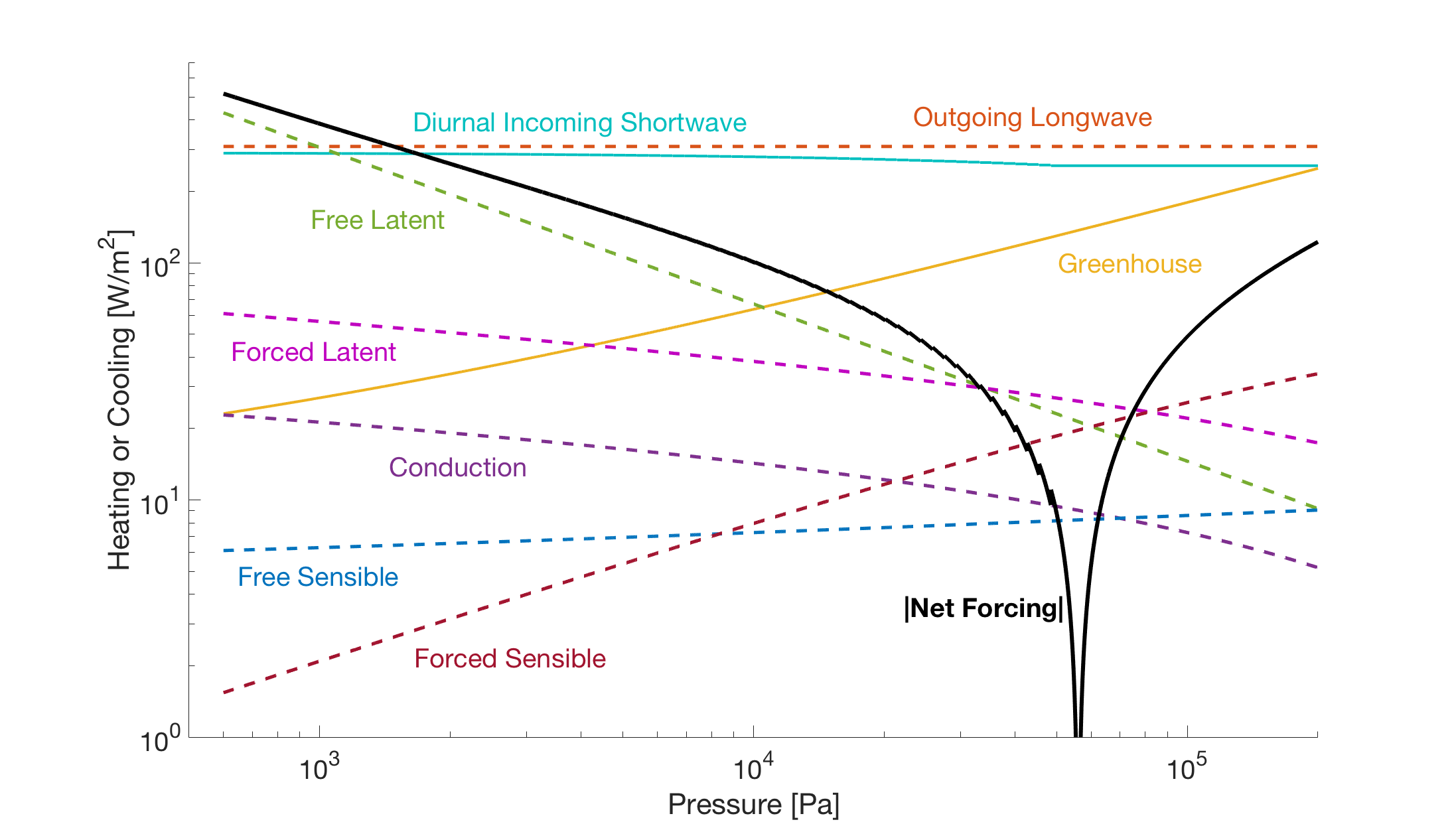}
\caption{\label{fig:pvsforce} Magnitude of each term in the energy balance versus pressure, at 77\% modern solar luminosity, which is the approximate solar luminosity at 3.5~Gya. Solid/dashed lines indicate warming/cooling. The lines show incoming shortwave solar radiation (light blue), outgoing longwave radiation (orange), greenhouse forcing (yellow), conductive cooling (purple), free latent cooling (green), forced latent cooling (magenta), free sensible cooling (dark blue), forced sensible cooling (dark red), and the magnitude of the net forcing (black). Small wiggles in the net forcing are artifacts of interpolation to calculate incoming shortwave radiation. The net forcing is negative for $P<56039$~Pa because latent cooling dominates over greenhouse forcing, and positive for $P>56039$~Pa.}
\end{figure}

\begin{figure}
\centering
\includegraphics[width=\linewidth]{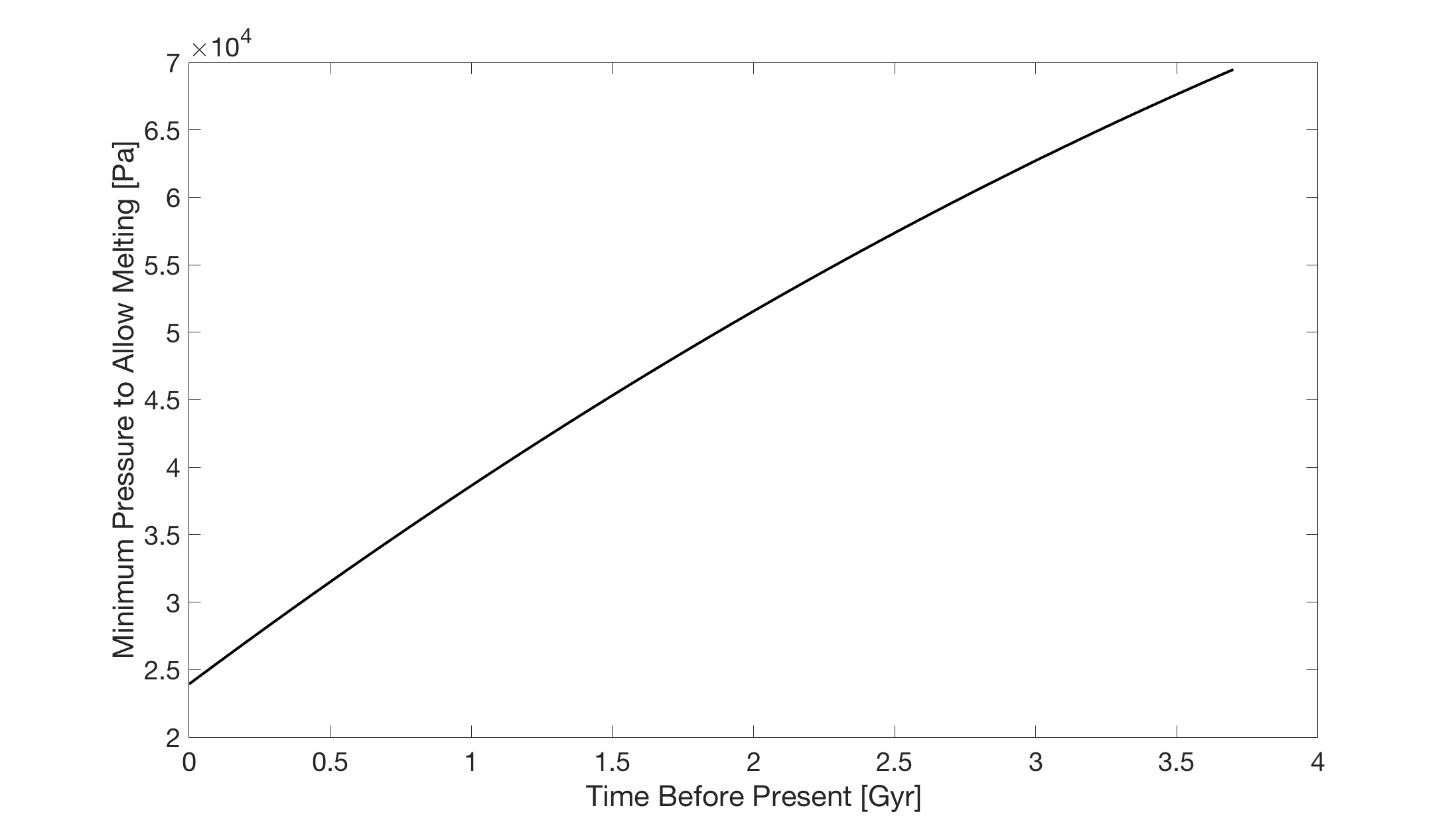}
\caption{\label{fig:minPovertime} Minimum pressure at which surface melting is permitted for a flat surface at $0\degree$ latitude as a function of time, assuming zero obliquity and values of eccentricity and longitude of perihelion equal to those for modern-day Mars. This curve was smoothed to remove artifacts of the interpolation to calculate incoming shortwave radiation. Earlier in time the minimum pressure to allow melting is higher because the Sun was fainter.}
\end{figure}

The final energy balance is given by $F_{net}=F_{SW}-F_{LW}+F_{gh}-F_{cond}-S_{fo}-S_{fr}-L_{fo}-L_{fr}$. A positive value of $F_{net}$ represented that there was excess energy available to melt snow on the surface, while a negative value of $F_{net}$ indicated that melting was prohibited. Figure \ref{fig:pvsforce} shows each of the terms in the energy balance model and the overall surface energy for a variety of atmospheric pressures at 77\% modern-day solar luminosity, which is approximately the solar luminosity at 3.5~Gya. At very low pressures, latent cooling dominates over most of the other terms and prevents melting. At higher pressures, the greenhouse forcing dominates over latent cooling, and melting occurs. This is because the greenhouse forcing scales as $\sqrt{P}$, while the free latent cooling scales as $P^{-2/3}$. Figure \ref{fig:minPovertime} shows the minimum pressure to permit melting as a function of time, assuming zero obliquity and values of eccentricity and longitude of perihelion equal to those for modern-day Mars. The minimum pressure to allow melting at 3.5~Ga is 0.676~bar. At earlier times, with the orbital parameters held constant, the lower solar luminosity means that a higher pressure is required in order to allow melting.

For each possible orbital history, an energy balance was computed at latitudes of $0\degree$, $20\degree$, and $40\degree$ in order to understand the relative importance of obliquity and atmospheric pressure changes on a variety of latitudes throughout the areas where young fluvial features are seen on Mars. Additionally, the water source regions we are modeling are on sloped surfaces with slopes of approximately $20\degree$, so energy balances were computed for a surface at $20\degree$ with a $20\degree$ pole-facing slope and for a surface at $40\degree$ with both a $20\degree$ pole-facing and $20\degree$ equator-facing slope.

\section{Results}
\label{sec:results}

\subsection{Relative Impact of Atmospheric Pressure and Obliquity}

\begin{figure}
\centering
\includegraphics[width=\linewidth]{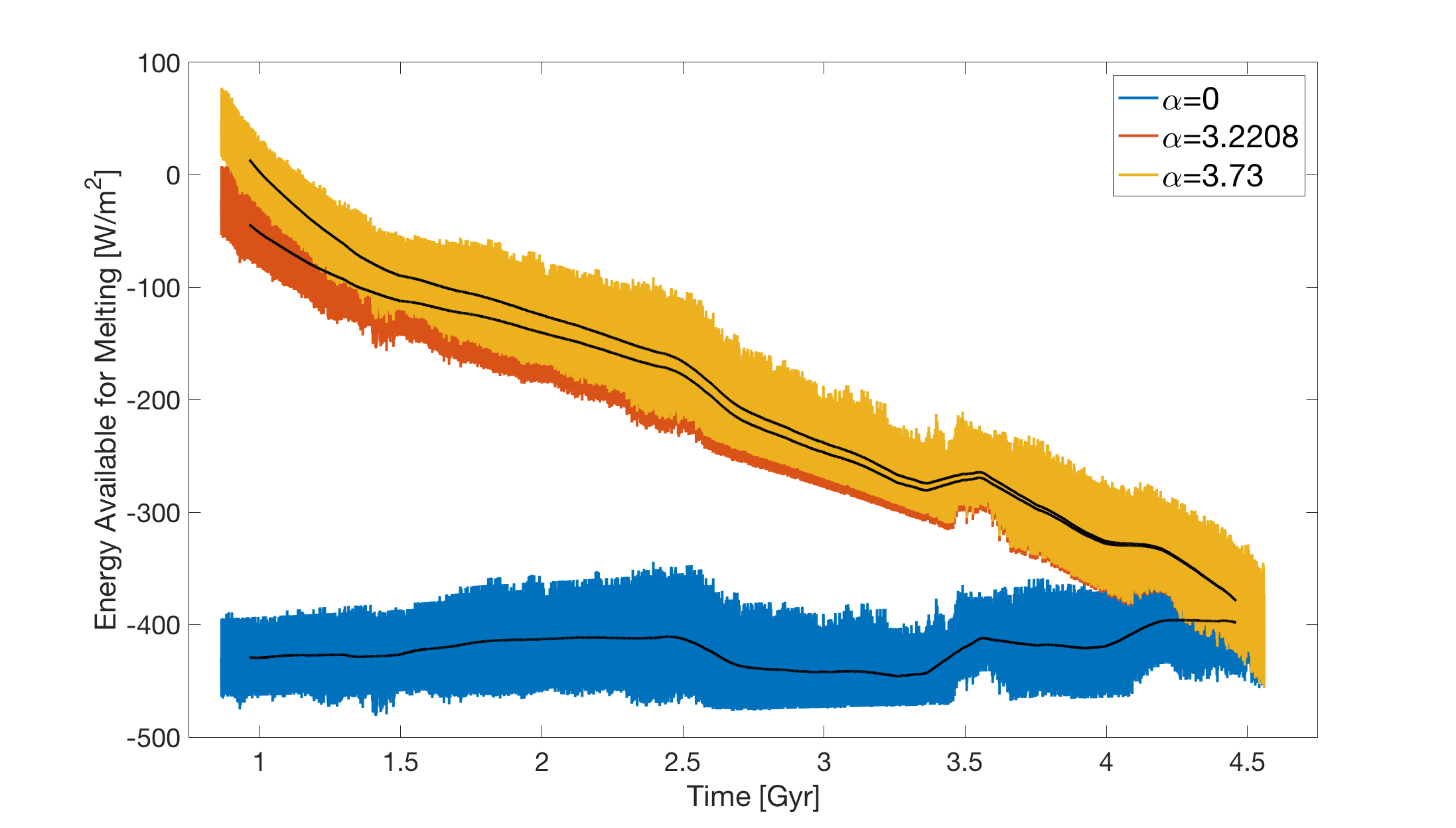}
\caption{\label{fig:ebalT23} Energy balance over time for obliquity track 1, for a surface at a latitude of $20\degree$ with a $20\degree$ pole-facing slope. The blue curve shows energy balance at constant pressure. The orange and yellow curves show energy balances for pressure loss parameterized by Equation \ref{eq:atmloss} with $\alpha=3.22$ and $\alpha=3.73$, respectively. $\alpha=3.22$ is the best estimate of $\alpha$ from initial MAVEN results, found by combining an estimate from MAVEN of the dependence of the loss rate on solar EUV flux with an estimate of the rate at which solar UV flux has changed over time \citep{Lillis2017,Tu2015}. $\alpha=3.73$ is 0.5~standard deviations away from the best estimate of $\alpha=3.22$ inferred from MAVEN data\citep{Lillis2017,Tu2015}. Black lines show 200-Myr averages, which average over orbital variations and show changes due to solar brightening and atmospheric loss. The increase in energy around 3.5~Gyr is where the obliquity rapidly changes from a mean obliquity around $10\degree$ to a mean obliquity around $40\degree$.}
\end{figure}

Figure \ref{fig:ebalT23} shows an energy balance history for a surface at $20\degree$ with a $20\degree$ pole-facing slope, using the obliquity history for track 1. The blue curve shows an energy balance at constant pressure, while the orange and yellow curves show energy balances with atmospheric loss parameterized by Equation \ref{eq:atmloss} with $\alpha=3.22$ ($P$=0.654~bars at 3.5~Gya) and $\alpha=3.73$ ($P$=1.133~bars at 3.5~Gya), respectively. Black lines show 200-Myr averages, which average over variations in eccentricity, obliquity, and longitude of perihelion. The upward trend that is visible in the earliest 1.5~Gyr of the fit to the blue curve demonstrates the effect of solar brightening. The downward trend of the fits to the orange and yellow curves show the effect of atmospheric loss. The bump in all curves around 3.5~Gyr shows the effect of a rapid change from a low mean obliquity around $10\degree$ to a high mean obliquity around $40\degree$, which increases the energy available for melting because a pole-facing surface at a latitude of $20\degree$ will receive more direct sunlight for an obliquity around $40\degree$ than for an obliquity around $10\degree$. For almost all times and values of $\alpha$, the energy balance is negative, indicating that melting is prohibited. In all potential orbital histories considered, only $\alpha \gtrapprox3.73$ allows melting, which indicates that atmospheric pressure has a stronger influence on whether melt is permitted at the surface than obliquity. However, this value of $\alpha$ is within $0.5\sigma$ of the value inferred from initial MAVEN data, indicating that large post-Noachian alluvial fans and deltas might be explained by a rate of atmospheric loss only slightly higher than that estimated by \citet{Lillis2017}.

\subsection{Comparison to Key Geologic Constraints}

For the high values of $\alpha$ that allowed melting, we compared the amount of melting and intermittency of melt periods to three key geologic constraints. First, the amount of fluvial sediment transported by large post-Noachian alluvial fans indicates that the longest runoff event must have had liquid water at the surface for $>10^{4}$~years \citep{Irwin2015}. Here, we use runoff event to mean a time interval during which seasonal meltwater runoff sustains lakes at an approximately constant lake level \citep{Irwin2015}. Second, the presence of olivine in the catchments of alluvial fans and deltas strongly suggests that the most recent runoff event must have lasted $<10^{7}$~years \citep{Stopar2006,Kite2017b}. Third, counting interbedded craters indicates that there must have been $>10^{7}$~years between the first and last runoff events \citep{Kite2017a}. Some estimates of melting in young channels, like those found in Lyot crater, suggest that the time between the first and last runoff events lasted for much longer than $10^{7}$~years \citep{Dickson2009}.

In order to produce melt, high temperatures are required, but the snow and ice also need to be available in the location where melting can occur. At obliquities greater than $40\degree$, snow and water ice are most stable near the equator, while for lower obliquities they are more stable at the poles \citep{Jakosky1985,Mischna2003,Forget2006}. If the amount of ice contained in the polar caps was spread over the equatorial region, it would produce an ice sheet about 100~m - 1~km thick, which would melt or sublimate in 100 - 10,000~years \citep{Madeleine2009}. This melting time is much less than the length of warm periods produced by obliquity changes in our model, and so the timescale for the ice to migrate to the coldest location on the planet is less than the timescale at which changes in obliquity shift the latitude of the coldest point on the planet. This means that when the obliquity is high, we can make the assumption that all of the snow and ice are found near the equator, where snow and ice are most stable. Therefore, for latitudes of $0\degree$ and $20\degree$, we assumed that melting could only have happened when the obliquity was greater than $40\degree$. For a latitude of $40\degree$, we assumed that snow and ice would only be available when the obliquity was between $30\degree$ and $40\degree$. Additionally, we assumed that runoff would not begin to occur until some amount of liquid water had pooled and refroze within the snow to create an impermeable ice layer for runoff \citep{Woo2012}, and so we only began recording a melt period when the energy available for melting went above 15~W/m$^{2}$.

\begin{figure}
\centering
\includegraphics[width=\linewidth]{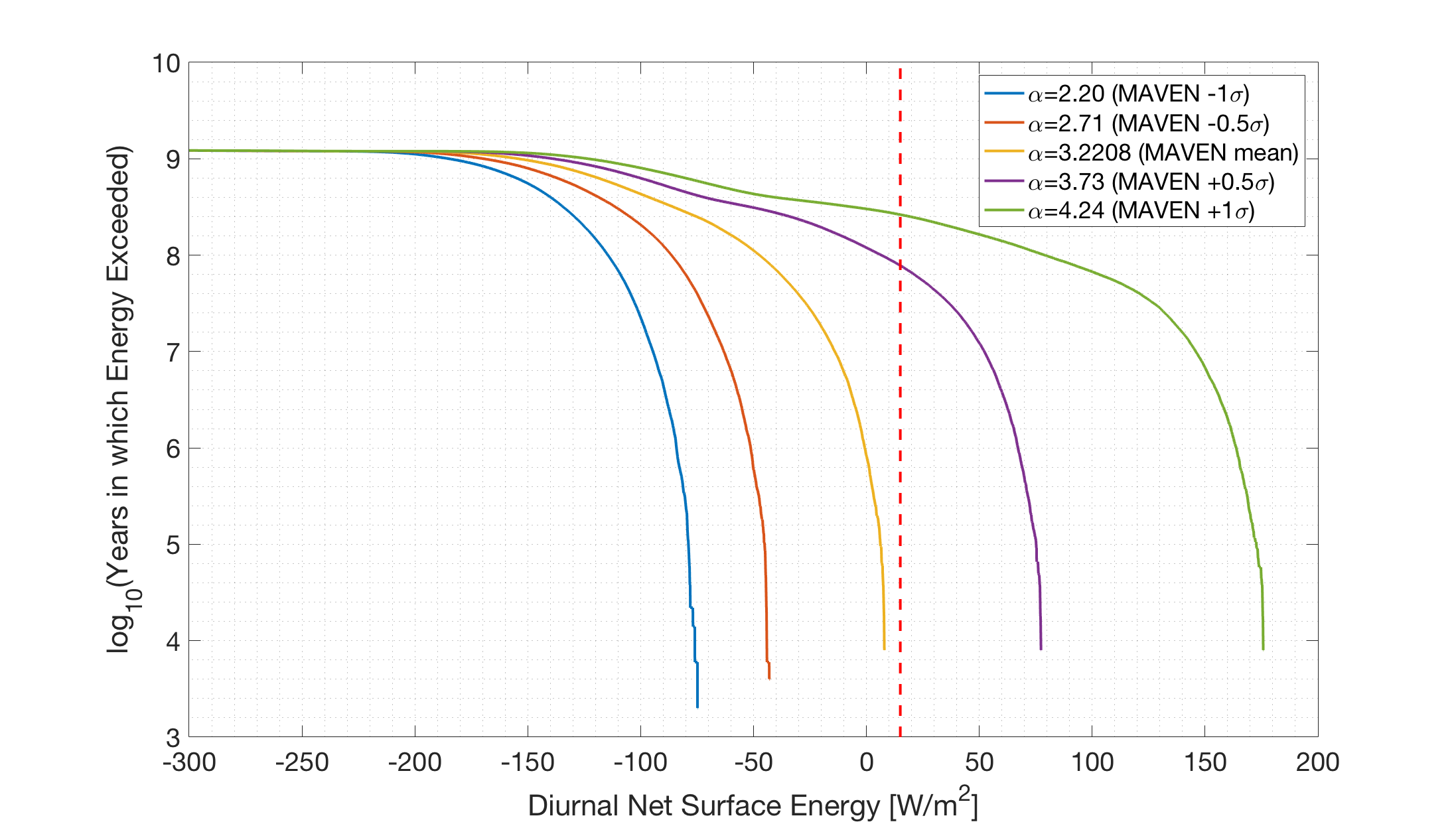}
\caption{Energy available for melting vs. the number of years that energy was exceeded for obliquity track 1 at $20\degree$ latitude with a $20\degree$ pole-facing slope, for $\alpha=2.20$, 2.71, 3.22, 3.73, and 4.24. For this graph, only times with obliquity $>40\degree$ are plotted, because snow and ice will only be available for melting at low latitudes during times of high obliquity \citep{Kite2013}. The red dashed line indicates a net energy of 15~W/m$^{2}$, above which runoff can occur. Only values of $\alpha$ above about 3.73 allowed any runoff.}
\label{fig:hist23}
\end{figure}

Figure \ref{fig:hist23} shows a plot of the net surface energy vs. the number of years that energy was exceeded for obliquity track 1. This plot shows results for a $20\degree$ pole-facing sloped surface at $20\degree$ latitude, with values of $\alpha$ within 1 standard deviation of the mean value inferred by \citet{Lillis2017}. Only years for which the obliquity was greater than $40\degree$ are counted, because these are the only years when ice would be available for melting. The vertical dashed red line indicates a surface energy of $15$~W/m$^{2}$, above which runoff would occur. For $\alpha=3.73$ ($P$=1.133~bars at 3.5~Gya) and $\alpha=4.24$ ($P$=2.020~bars at 3.5~Gya), some runoff is produced when the orbital history is at high obliquity. For $\alpha=3.73$, the longest melting event lasts ${4.44 \times 10^{7}}$~years, the most recent melting event lasts ${2.40 \times 10^{4}}$~years, and the total time between the first and last melting events is ${2.78 \times 10^{8}}$~years. For $\alpha=4.24$, these values are ${2.55 \times 10^{8}}$~years, ${1.80 \times 10^{4}}$~years, and ${5.10 \times 10^{8}}$~years, respectively. Therefore, both of these values of $\alpha$ produce melt periods that match the three geologic constraints. For $\alpha=3.73$, the rate of runoff production is approximately 0.05~mm/hr for approximately 18~Myr, which is consistent with estimated runoff production rates of 0.03-0.4~mm/hr \citep{Irwin2015,Morgan2014,Dietrich2017}.

Smaller values of $\alpha$, corresponding to lower past pressures, do not allow melting under the current model, but this model only considered greenhouse forcing from a pure CO$_{2}$ atmosphere. Some amount of non-CO$_{2}$ greenhouse forcing could increase the surface energy balance enough that lower values of $\alpha$ would still produce some melting, as discussed in Section \ref{sec:future}.

\begin{figure}
\centering
\includegraphics[width=\linewidth]{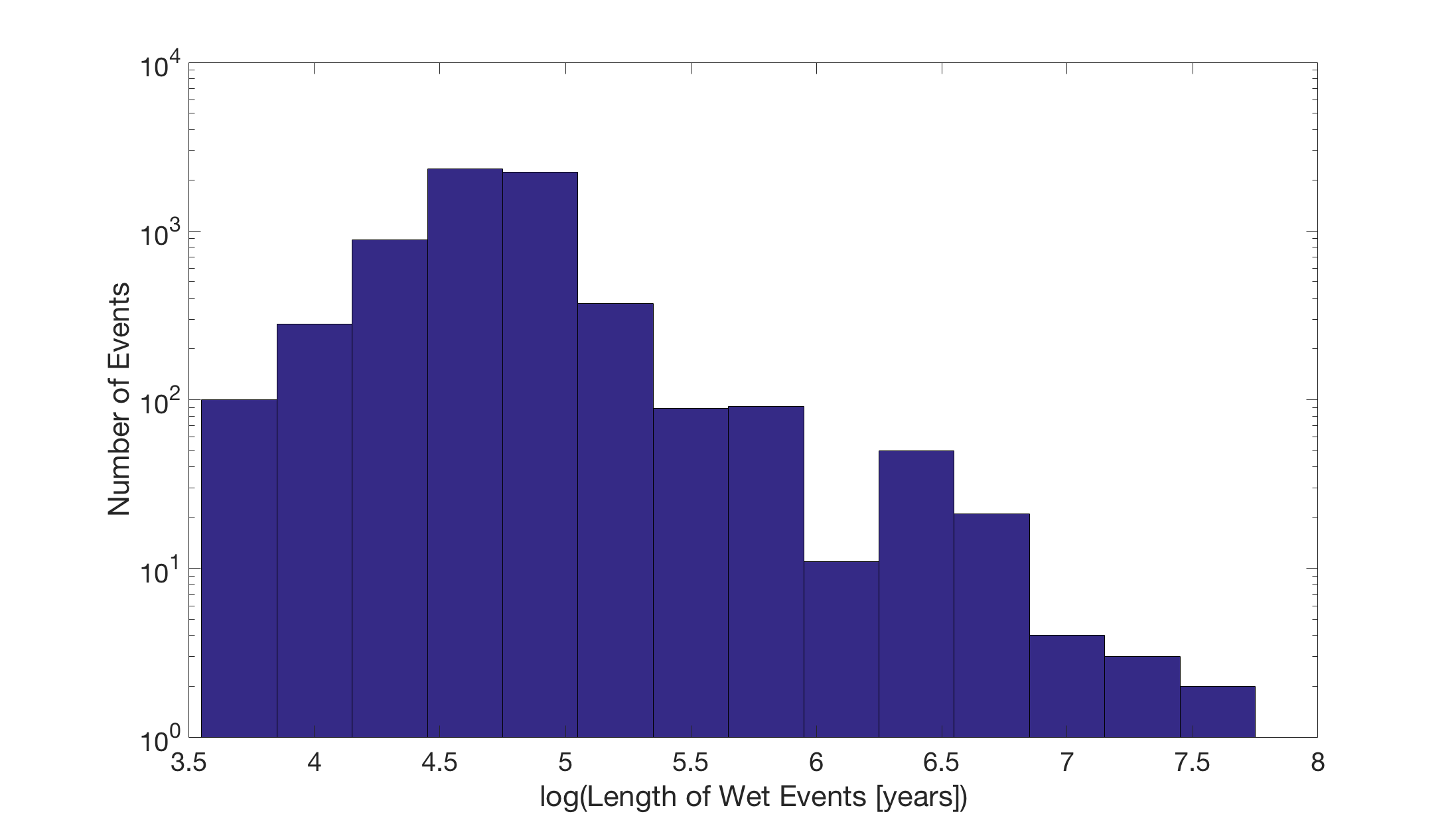}
\caption{Histogram of the length of periods in which melting is allowed for obliquity track 1 for all latitudes and slopes considered, and for $\alpha=3.73$. In all cases, we assume that a melt period begins when the energy balance goes above 15~W/m$^{2}$ and ends when it goes below 0~W/m$^{2}$. We also assume melting can only occur at low latitudes for obliquities $>40\degree$ and at higher latitudes for obliquities between $30\degree$ and $40\degree$. See text for discussion.}
\label{fig:histwet}
\end{figure}

The distribution of the length of melting periods shows how the melt conditions depend on orbital cycles. Figure \ref{fig:histwet} shows a histogram of the length of melting periods in obliquity track 1 for all latitudes and slopes considered for $\alpha=3.73$. During all of these melting periods, melting occurs every year during the warmest part of the year. The majority of time periods when melting is allowed last around 100,000~years, which is about the timescale of a single peak-to-trough oscillation of obliquity, eccentricity, or longitude of perihelion. Therefore, most of the melting occurs in short time periods about 100~kyr long during optimal orbital conditions. For example, at high latitudes much of the melting occurs when the eccentricity and longitude of perihelion line up such that Mars is near perihelion at solstice and the obliquity is such that the amount of direct sunlight is maximized during the warmest part of the day. There are fewer melting periods of 1 to 10~Myr duration, which correspond to the length of larger oscillations in obliquity and eccentricity. The couple of periods that last longer than 50~Myr corresponds to long periods of optimal conditions due to a much higher mean obliquity. Between these intermittent melting periods were several dry periods, which could affect the ability of organisms to survive on the surface of Mars. All eight obliquity tracks contained at least one such dry period longer than 10~Myr, which is long enough that the top few meters of soil would be sterilized by galactic cosmic radiation even for radiation-resistant microbes like \textit{D. radiodurans} \citep{Hassler2014}.

\section{Discussion}
\label{sec:discussion}

\subsection{Model Assumptions}
\label{sec:assume}

Our model made several assumptions about physical characteristics of Mars's surface and atmosphere which made it more likely to produce liquid water. Here we consider the key assumptions and how changes to these assumptions could affect our conclusions. First, the surface albedo was assumed to be 0.3. Assuming a higher albedo value would reduce the net energy flux into the surface, and so less melting would be permitted. Second, we assumed a thermal inertia of $\approx$275~J m$^{-2}$ K$^{-1}$ s$^{-1/2}$. A higher thermal inertia value would also reduce the time over which melting was permitted for a given value of $\alpha$ because the surface would conductively cool faster. Third, we assumed no refreezing or infiltration losses of meltwater, both of which would decrease the amount of runoff produced. These assumptions combine to create the most optimal conditions for producing runoff.

We also assumed the latitudes where ice is stable change abruptly at obliquities of $40\degree$ and $30\degree$. In other words, we assumed that at an obliquity just below $40\degree$, all ice would be located at middle or high latitudes, and at an obliquity just above $40\degree$, all ice would be in the equatorial region. In reality, this transition may be less abrupt. However, since we found that atmospheric pressure was the primary control on melting, the exact value of obliquity at which ice migrates to the equator should not have a strong influence on the results. Additionally, it is possible that, at high pressures, water ice will be located near the equator for all obliquities \citep{Wordsworth2016}. If this were the case, melting at the equator would potentially occur even at lower obliquities, because water ice would still be in the equatorial region.

\subsection{Open Questions}
\label{sec:future}

We found that a rate of atmospheric loss consistent with that inferred by \citet{Lillis2017} to within 0.5 standard deviations allowed melting and produced an amount of runoff which matched three key geologic constraints \citep{Lillis2017,Tu2015,Irwin2015,Stopar2006,Kite2017b,Kite2017a}. However, atmospheric loss by escape to space is not the only process that may have affected the energy balance on Mars's surface. Non-CO$_{2}$ forcing from a variety of sources could have permitted melting at lower atmospheric pressures. Our model did not include clouds, which can warm the surface \citep{Ramirez2017}. At higher obliquities, when water ice is most stable in the equatorial region, the warm temperatures at those low latitudes might produce more water vapor, which could form into cirrus clouds that would warm the planet. The water vapor would also provide direct vapor warming on the order of a few Kelvin \citep{Mischna2013}. We also did not consider warming from other minor atmospheric constituents. In particular, CO$_{2}$-H$_{2}$ and CO$_{2}$-CH$_{4}$ collision-induced absorption (CIA) can have a significant impact on the greenhouse forcing \citep{Wordsworth2017}. Because melting only occurs before 3.3~Gya in our model, additional warming may also be necessary to explain observations of anomalously young ($<2$~Gya) supraglacial channels at the low-elevation Lyot crater \citep{Dickson2009} and in isolated cases elsewhere \citep{Fassett2010}. Some very young alluvial fans, like those in the 5-Myr old Mojave crater, are likely related to impact processes \citep{WilliamsR2008}.

Another possible explanation for these young fluvial features is higher atmospheric pressure than was estimated in our model. The atmospheric pressure could have been higher in the past if CO$_{2}$ was lost through other mechanisms, such as chemical fixation in deep aquifers \citep{Chassefiere2011} or basal melting of a CO$_{2}$ ice cap \citep{Kurahashi2006,Longhi2006}. Higher atmospheric pressure would increase the amount of runoff compared to our calculations. Additionally, volcanic degassing could increase the amount of melt by adding more CO$_{2}$ (and other greenhouse gases such as CH$_{4}$, SO$_{2}$, and H$_{2}$S) into the atmosphere. However, \citet{Stanley2011} calculate that Mars magmas have less CO$_{2}$ than on Earth because the Mars mantle is more reducing. \citet{Stanley2011} estimate that the volcanic CO$_{2}$ content is approximately 50-70~ppm, which would correspond to about 30~mbar of volcanic outgassing. If the Martian atmosphere was more oxidizing, the volcanic CO$_{2}$ content could be as high as 500-700~ppm, but the pressure contribution would still be relatively small, about 300~mbar \citep{Stanley2011}. These \citet{Stanley2011} numbers imply a substantial downward revision from \citet{Craddock2009}, who estimated that Martian volcanoes contained 0.7~wt.\%~CO$_{2}$. Carbonate formation may also have been underestimated, which would mean higher past atmospheric pressures could exist without invoking larger loss rates due to solar UV flux \citep{Hu2015}.

\section{Conclusions}
\label{sec:conclusions}

We wrote a minimal zero-dimensional energy balance model containing only effects of a CO$_{2}$ atmosphere, solar brightening, and obliquity changes, to investigate whether conditions on the surface of Mars could have produced enough melting to explain observed large ($>10 \text{ km}^{2}$) alluvial fan features in the last 3.5~Gyr. We find that melting is only permitted for thick atmospheres that require high rates of atmospheric loss, and that the effect of high atmospheric pressure is more important than varying values of obliquity in producing melt over the history of Mars. However, if we make the end-member assumption that all atmospheric loss is due to escape to space and all escaping oxygen is from CO$_{2}$, rates of atmospheric loss consistent to within 0.5~standard deviations with the best estimate of atmospheric loss based on initial results from the MAVEN mission produce melting. Additionally, the amount and timing of runoff from this melting matches three specific, recently-obtained key geologic constraints on the formation of alluvial fans \citep{Lillis2017,Tu2015,Irwin2015,Stopar2006,Kite2017b,Kite2017a}.

Although our model only produced melting for very high atmospheric pressures, additional warming due to, for example, cirrus clouds, water vapor warming, or collision-induced absorption could warm the surface enough to allow melting at lower pressures \citep{Ramirez2017,Mischna2013,Wordsworth2017}. Further research into non-CO$_{2}$ warming can determine whether it produces a large enough effect to allow melting at lower past atmospheric pressures.

\begin{acknowledgments} The authors would like to thank Rob Lillis, Itay Halevy, John Armstrong, and the University of Chicago Research Computing Center. The authors would like to thank the referees, whose comments greatly improved the paper. This paper contains no new data, but the energy balance model is available at https://github.com/meganmansfield/MarsEnergyBalance. We are grateful for support from NASA grant NNX16AG55G.
\end{acknowledgments}


\end{document}